\documentclass[aps,pra,superscriptaddress,twocolumn,nofootinbib,showpacs,floatfix]{revtex4-1}

\usepackage{graphicx}
\usepackage{dcolumn}
\usepackage{bm}
\usepackage{color}
\usepackage[colorlinks=true,urlcolor=blue,citecolor=blue, linkcolor = blue]{hyperref}
\usepackage{amssymb,ulem,amsmath}
\def\be{\begin{equation}}
\def\ee{\end{equation}}
\def\ba{\begin{array}{lll}}
\def\ea{\end{array}}
\def\ber{\begin{eqnarray}}
\def\eer{\end{eqnarray}}

\newcommand{\braket}[2]{\left\langle #1 | #2 \right\rangle}
\newcommand{\qql}{\textquotedblleft}
\newcommand{\qqr}{\textquotedblright}

\begin{document}
\title{Quantum Shock Waves and Population Inversion in Collisions of Ultracold Atomic Clouds}

\author{Sebastiano Peotta}
\email{speotta@physics.ucsd.edu}
\author{Massimiliano Di Ventra}
\affiliation{Department of Physics, University of California, San Diego, California 92093, USA}
\begin{abstract}
Using Time-Dependent Density Matrix Renormalization Group (TDMRG) we study the collision of one-dimensional atomic clouds confined in a harmonic trap and evolving with the Lieb-Liniger Hamiltonian. It is observed that the motion is essentially periodic with the clouds bouncing elastically, at least on the time scale of the first few oscillations that can  be resolved with high accuracy. This is in agreement with the results of the \qql quantum Newton cradle\qqr  experiment of Kinoshita \textit{et al.} [Nature \textbf{440}, 900 (2006)]. We compare the results for the density profile against a hydrodynamic description, or generalized nonlinear Schr\"odinger equation, with the pressure term taken from the Bethe Ansatz solution of the Lieb-Liniger model. We find that hydrodynamics can describe the breathing mode of a harmonically trapped cloud for arbitrary long times while it breaks down almost immediately for the collision of two clouds due to the formation of shock waves (gradient catastrophe). In the case of the clouds' collision TDMRG alone allows to extract the oscillation period which is found to be measurably different from the breathing mode period.
Concomitantly with the shock waves formation we observe a local energy distribution typical of population inversion, \textit{i.e.}, an effective negative temperature.
Our results are an important step towards understanding the hydrodynamics of quantum many-body systems out of equilibrium and the role of integrability in their dynamics.

\end{abstract}
\maketitle

\section{Introduction}\label{sec:intro}
The quantum dynamics of closed many-body  quantum systems is relatively unexplored~\cite{Polkovnikov:2011} and has become the subject of active research only recently with the advent of highly tunable ultracold atomic gases~\cite{Bloch:2008,Lewenstein:2007}. In these systems the almost perfect decoupling from the external environment and the long time scales allow to study  details of the quantum dynamics that are not easily accessible, e.g., in solid-state systems.
An example of the tunability of ultracold gases is the use of  optical lattices~\citep{Bloch:2008} to freeze the transverse motion  and confine the gas in one dimension, a regime where quantum fluctuations play a prominent role~\citep{Giamarchi_book}. Interestingly several 1D  Hamiltonians  relevant to ultracold gases are known to be integrable, \textit{i.e.}, they possess an infinite number of local conserved quantities~\cite{Sutherland_book,Korepin_book,Cazalilla:2011}.

The implications of integrability on the time evolution of a quantum system is far from being understood as shown in the highly debated  \qql quantum Newton cradle\qqr experiment of Ref.~\onlinecite{Kinoshita:2006} which promptly followed the realization of a Tonks-Girardeau gas~\citep{Girardeau:1960,Girardeau:1965,Kinoshita:2004,Tolra:2004,Paredes:2004,Kinoshita:2005}. In this latter work  a 1D gas of bosons interacting via a contact potential (Lieb-Liniger model~\cite{Lieb:1963a,Lieb:1963b}) is separated in two symmetric clouds that subsequently collide in a harmonic trap.
Interestingly the clouds bounce off each other several hundred times without noticeable decay of the oscillatory motion. On the other hand, in three dimensions the dynamics are dramatically different with the clouds merging in a single motionless lump - \textit{i.e.}, a thermal state - after few bounces~\cite{Kinoshita:2006}.
This sharp difference between the behaviors in one and three dimensions has triggered a vast amount of theoretical work~\cite{Rigol:2013,Collura:2013} aimed at understanding the nature of the asymptotic state (if any) reached when the time evolution is dictated by an integrable Hamiltonian.

It is fairly clear that integrability manifests itself only in out of equilibrium dynamics whose accurate description requires eigenstates with an energy substantially larger than the ground state energy~\cite{Kinoshita:2006,Ronzheimer:2013,Arzamasovs:2013}. On the contrary 1D integrable and nonintegrable models behave alike when the dynamics are restricted to the low energy spectrum, \textit{i.e.} in the linear response regime. For instance, a large class of integrable and nonintegrable 1D Hamilonians are known to fall in the universality class of the Tomonaga-Luttinger liquid, an integrable model~\cite{Giamarchi_book,Cazalilla:2011,Gogolin_book,Stone_book}.

As illustrated  in Ref.~\onlinecite{Kinoshita:2006}, ultracold gases can be easily driven in different out of equilibrium and nonlinear regimes while a substantial effort is needed to probe only their linear response~\cite{Bloch:2012}. Unfortunately not many approaches are available to study the out of equilibrium dynamics of interacting quantum systems. Only in 1D the Time-Dependent Density Matrix Renormalization Group (TDMRG)~\cite{White:2004,Vidal:2004,Daley:2004,tdmrg} allows the -- numerically exact -- simulation of the real time dynamics for arbitrary Hamiltonians, with the restriction that the entanglement content of the evolving wavefunction is initially not too large and not growing too rapidly in time~\cite{tdmrg}.

Alternatively one can discard the fine-grained description of a system, such as the full wavefunction (or an approximation thereof), and focus directly on the collective dynamics of the observables of interests, the particle density being the easier to access in the context of quantum gases. The collective field approach has been very successful for Bose-Einstein condensates, called also coherent matter waves~\cite{Dalfovo:1999}. At much lower temperatures than the condensation temperature the only relevant degree of freedom of a gas of weakly interacting bosons is a space-dependent complex order parameter, namely the wavefunction in which a  macroscopically large number of particles condense, and interactions can be safely accounted for at the mean-field level~\cite{Castin:2001}. The evolution of the complex order parameter is governed by the celebrated Gross-Pitaevskii equation~\cite{Gross:1961,Gross:1963,Pitaevskii:1961} which has proved to be very effective in providing quantitative predictions for the static, dynamic, and thermodynamic properties of trapped Bose gases~\cite{Dalfovo:1999}.

The Gross-Pitaevskii equation is equivalent to the standard Euler's equations of fluid dynamics for an inviscid fluid, albeit with an additional \qql quantum pressure\qqr term.
The Tomonaga-Luttinger theory of 1D many-body systems is sometimes called \qql hydrodynamics\qqr~\cite{Arzamasovs:2013} or \qql harmonic fluid\qqr approach~\cite{Haldane:1981a,Haldane:1981b} since the canonical fields in the Hamiltonian are the integrated density $\phi = \int \rho $ and velocity $\theta = \int v$ and represent the relevant collective modes at low energies and long wavelength~\cite{Giamarchi_book}. While the Gross-Pitaevskii's is in essence a classical nonlinear hydrodynamics, the Tomonaga-Luttinger model is a linear -- noninteracting -- quantum field theory. Nonlinear extensions of the Tomonaga-Luttinger theory have been discussed in several contexts~\cite{Bettelheim:2006a,Bettelheim:2006b,Bettelheim:2007,
Bettelheim:2008,Schmidt:2009,Schmidt:2010,Imambekov:2012}, but throughout this work  \qql hydrodynamics description\qqr will stand for a system of nonlinear equations for a classical fluid. Incidentally, this is the same approach used in Time-Dependent Density Functional Theory (TD-DFT)~\cite{Gross:1984} in particular in its orbital free formulation~\cite{Ligneres:2005}, where the density is the sole dynamical variable.

For experiments such as the collision between degenerate  clouds comprising a large number of interacting atoms a collective field description is usually the only available option. Various phenomena have been studied in collision experiments, such as the interference of matter waves~\cite{Ketterle:1997}, dispersive shock waves in BECs~\cite{Hoefer:2006,Meppelink:2009}, superfluidity, shock wave formation and domain wall propagation in the unitary Fermi gas~\cite{Joseph:2011,Bulgac:2012,Salasnich:2012,Ancilotto:2012}, spin transport~\cite{Sommer:2011}, and lack of thermalization in quasi-integrable 1D systems~\cite{Kinoshita:2006,Polkovnikov:2011}.

With reference to the experimental setup of Ref.~\onlinecite{Kinoshita:2006}, we study the collision of two clouds of one dimensional bosons for arbitrary interaction strength, by means of a Time-Dependent Density Matrix Renormalization Group (TDMRG) approach~\cite{White:2004,Vidal:2004,Daley:2004,tdmrg} based on a Matrix Product State (MPS) approximation of the full wavefunction.

A first important result presented here is that the numerical simulation of the experiment in Ref.~\onlinecite{Kinoshita:2006} for the first few ($\sim 3$) oscillations is within reach of TDMRG and we provide details on how this has been accomplished. Moreover, if the time evolution is computed accurately, the entanglement is slowly growing in the quenches that we perform, a fact that can possibly allow to reach times much longer than the ones considered in this work. Assessing the maximal evolution time allowed by TDMRG requires a more accurate analysis of the numerical errors which is beyond the scope of the present work. Therefore the important questions of thermalization and of the nature of the asymptotic state is not the focus here. However we put forward a definition of local temperature that could be useful in this context (see below).

A second result presented here is the accurate comparison of the exact quantum dynamics with a generalized Gross-Pitaevskii equation or \textit{generalized nonlinear Schr\"odinger equation} (GNLSE)~\cite{Korepin_book} which, in hydrodynamic form, contains a pressure term derived from the Bethe Ansatz solution of the Lieb-Liniger model. This is the best available hydrodynamic description for the present problem. While hydrodynamics works for several oscillations for the breathing mode, in the case of the clouds' collision the formation of shock waves leads to a chaotic behavior which is not reflected in the periodic behavior shown by the TDMRG data. Only from the latter the oscillation period as a function of the interaction strength can be accurately extracted and it is found to be different from the breathing period, an easily testable prediction. This result emphasizes that a better understanding of quantum shock waves is instrumental to a -- at least qualitatively -- correct hydrodynamic description of 1D quantum gases.

Finally, we further characterize the formation of shock waves by studying the
Wigner distribution function, a tool used by other authors in the context of shock wave dynamics of free fermions~\cite{Bettelheim:2012,Mirlin:2012}.
Starting  from the Wigner function, we show how it is possible to define a \textit{local energy distribution function} and that at the onset of shock waves formation the latter shows \textit{population inversion}, \textit{i.e.} higher energy states are more occupied than lower energy ones. Recently~\cite{Braun:2013} a \textit{negative temperature}, namely a population inversion in the energy distribution of the motional degrees of freedom of atomic gases has been realized. Moreover, it was shown in Ref.~\onlinecite{Braun:2013} that population inversion does not necessarily imply a fast decay to the true thermal equilibrium state, thus showing the quite unique properties these systems possess. We suggest that the small thermalization rate and absence of visible decay of the oscillatory motion in the density profiles observed both in Ref.~\onlinecite{Kinoshita:2006} and in our simulations are a dynamical manifestation of the same remarkable (meta-)stability of the negative temperature state realized in Ref.~\onlinecite{Braun:2013}. In fact, we employ a possible definition of local temperature out of equilibrium -- put forward in Ref.~\onlinecite{DiVentra:2009} -- and find again negative values in correspondence of the shock wave formation time.

\section{Model and  methods}
\subsection{\label{sec:quenches} Lieb-Liniger model and quenches}
The Lieb-Liniger model~\cite{Lieb:1963a,Lieb:1963b} provides an excellent description of one dimensional ultracold bosonic atoms~\cite{Olshanii:1998,Dunjko:2001}. In terms of a bosonic field $\hat \Psi(x)$ its Hamiltonian reads
%
%
\begin{equation}\label{eq:ll}
\mathcal{\hat H}_{\rm LL} =  \int dx\, \bigg[\frac{\hbar^2}{2m}|\partial_x \hat \Psi(x)|^2 + \frac{g_B}{2}|\hat \Psi(x))|^4 +V(x)|\hat \Psi(x)|^2\bigg],
\end{equation}
where $g_B \in \left[0,+\infty\right]$ is a coupling constant and $m$ is the atom mass. In the following we will consider a time-dependent external potential $V(x,t)$ changing abruptly at $t=0$ (\textit{quench}).
Hamiltonian~(\ref{eq:ll}) is integrable for any $g_B$ when $V(x)=0$, while for $V(x)\neq0$ the exact eigenstates and eigenvalues are known for free bosons $g_B=0$ and hard-core bosons $g_B = +\infty$, the latter being equivalent to free fermions according to the Bose-Fermi mapping theorem~\cite{Girardeau:1960,Girardeau:1965}.

We consider two kinds of quench. In the first one the external potential is an harmonic well
with a sudden change in frequency at $t=0$
\begin{equation}\label{eq:post_pot}
V(x,t) = \frac{1}{2}m\omega^2_1(t)x^2,\quad \omega_1(t>0)
= \frac{\omega_1(t\leq0)}{\sqrt{3}}\,.
\end{equation}
This excites the \textit{breathing mode} of a gas initially in the ground state.
In a second kind of quench we prepare the gas in the ground state of the potential
\begin{equation}\label{eq:init_pot}
V(x,t \leq 0) = \frac{1}{2}m\omega^2_0\frac{(x^2-D^2)^2}{4D^2}\,,
\end{equation}
in order to have two clouds of particles separated by a distance $\sim D$,
and we let it evolve for $t>0$ in the harmonic potential~(\ref{eq:post_pot}) with $\omega_1 < \omega_0$ (\textit{microcanonical picture of transport}~\cite{mybook,chien}).
The values of the frequencies $\omega_0$ and $\omega_1$ depend on the interaction strength and are reported in Table~\ref{tb:frequencies}.

\subsection{\label{sec:tdmrg}TDMRG simulations}
It is possible to access the dynamics of~(\ref{eq:ll}),~(\ref{eq:post_pot}) and~(\ref{eq:init_pot}) in an essentially exact fashion using TDMRG. TDMRG has been applied mainly to lattice systems for relatively short time scales~\cite{Kollath:2005}, but simulations of systems in the continuum limit and for quite long time scales (of the order of several periods) are feasible~\cite{Muth:2010,Muth:2011,Peotta:2011,Peotta:2012,Knap:2013,
White:2012,Caux:2012}. Details are provided in Appendix~\ref{appendix:tdmrg}. In essence we use a wavefunction in MPS form that explicitly conserves the number of particles and is evolved in time using a sixth order Trotter expansion \cite{tdmrg,Peotta:2011,Muth:2011,Peotta:2012}. Moreover the sizes of the MPS matrices are allowed to change dynamically both in space and in time by fixing the discarded weight [see App.~\ref{appendix:tdmrg}].
 In our simulations we have employed two different discretizations~ of~(\ref{eq:ll})~\cite{Muth:2010}, either as a Bose-Hubbard model (nonintegrable), or as a XXZ spin chain (integrable) using the Bose-Fermi mapping for arbitrary interaction strength $g_B$~\cite{Muth:2010,Cheon:1999,Cheon:1998}. The lattice Hamiltonians are given in Appendix~\ref{appendix:discretization}. No substantial difference in the results has been observed.
Importantly we find that, for an accurate time evolution of the MPS, the entanglement entropy is bounded  or very slowly increasing~\cite{SOM} which translates in a manageable size of the MPS. Thus, in principle, longer times could be explored but we show only results for the first few half periods $\tau = \pi/\omega_1(t>0)$.

In the following, lengths are expressed in units of the lattice spacing $a$ of the discretized model, which is a small, but otherwise arbitrary length scale, energy is in units of $J = \hbar^2/(2ma^2)$, and time in units of the post-quench oscillation half period $\tau = \pi/\omega_1(t>0)$. The interaction parameter $g_B$ is given in units of $Ja$. Occasionally we use the Lieb-Liniger parameter $\gamma = mg_B/(\hbar^2\rho) $ with $\rho = 0.05/a$, an indicative value of the density in the inhomogeneous system considered here.
We ensured, by separately tuning $\omega_0$ and $\omega_1$ for each value of the coupling $g_B$ [see Table~\ref{tb:frequencies}], that the particle density \textit{per site} never exceeds $\sim0.15$, thus lattice effects are negligible (continuum limit)~\cite{Peotta:2011,Peotta:2012,Muth:2011}.

In our simulation we consider $N =20$ particles in a lattice of length $L=600a$. The clouds' distance is fixed at $D = 120a$ [see~(\ref{eq:init_pot})]. With this choice of distance the two clouds are always partially overlapping while simulations for well separated clouds are more numerically demanding.

In the actual experiment~\cite{Kinoshita:2006} the number of particles varies between $40$ and $250$, figures not far from the one used in our simulations. Moreover we will see that the dynamics can be  well described in the local density approximation, namely using the local pressure calculated in the thermodynamic limit. Thus our results are significant for much larger system sizes, a fact that we explicitly verified in the case of free fermions where a scaling in the number of particles can be easily performed. \

\begin{table}
\begin{tabular}{c|c|c|c}
$g_B/(Ja)$  &  $\hbar \omega_0/J$  & $\hbar\omega_1/J$ & $\tau  J/ \hbar$\\ \hline
0.0 		&  0.0005  		 &  0.0003 		 	 &  11107		     \\
0.002	& 0.0005			 & 0.0003			 &  11107		     \\
0.02		& 0.0009			 & 0.0006			 &   4967		     \\
0.2		& 0.0025			 &  0.0012			 &   2618		     \\
0.6		& 0.0040			 &  0.0017			 &   1756  		     \\
1.0		& 0.0046			 &  0.0022			 &   1433		     \\
1.4		& 0.0049			 &  0.0024			 &   1328		     \\
2.0		& 0.0055			 &  0.0025			 &   1258		     \\
$+\infty$ & 0.0063       &  0.0033            &   956
\end{tabular}
\caption{\label{tb:frequencies} Frequencies $\omega_0$ [Eq.~(\ref{eq:init_pot})] and $\omega_1$ [Eq.~(\ref{eq:post_pot})] and oscillation half period $\tau=\pi/\omega_1(t>0)$ used in the simulations according to the value of $g_B$. The table refers to the collision quench. In the case of the quench exciting the breathing mode the initial trapping frequency $\omega_1(t\leq 0)$ is given in the third column of the table, and the post-quench frequency by $\omega_1(t>0) = \omega_1(t\leq 0)/\sqrt{3}$ [Eq.~(\ref{eq:post_pot})]. Times are in units of $\hbar/J= ma^2/\hbar$. Changing these parameter with the interaction strength is important in order to keep the on-site density roughly constant when the compressibility of the gas varies. The above choice works well as one can see in Fig.~\ref{fig1}.}
\end{table}

\begin{figure*}[t]
\includegraphics{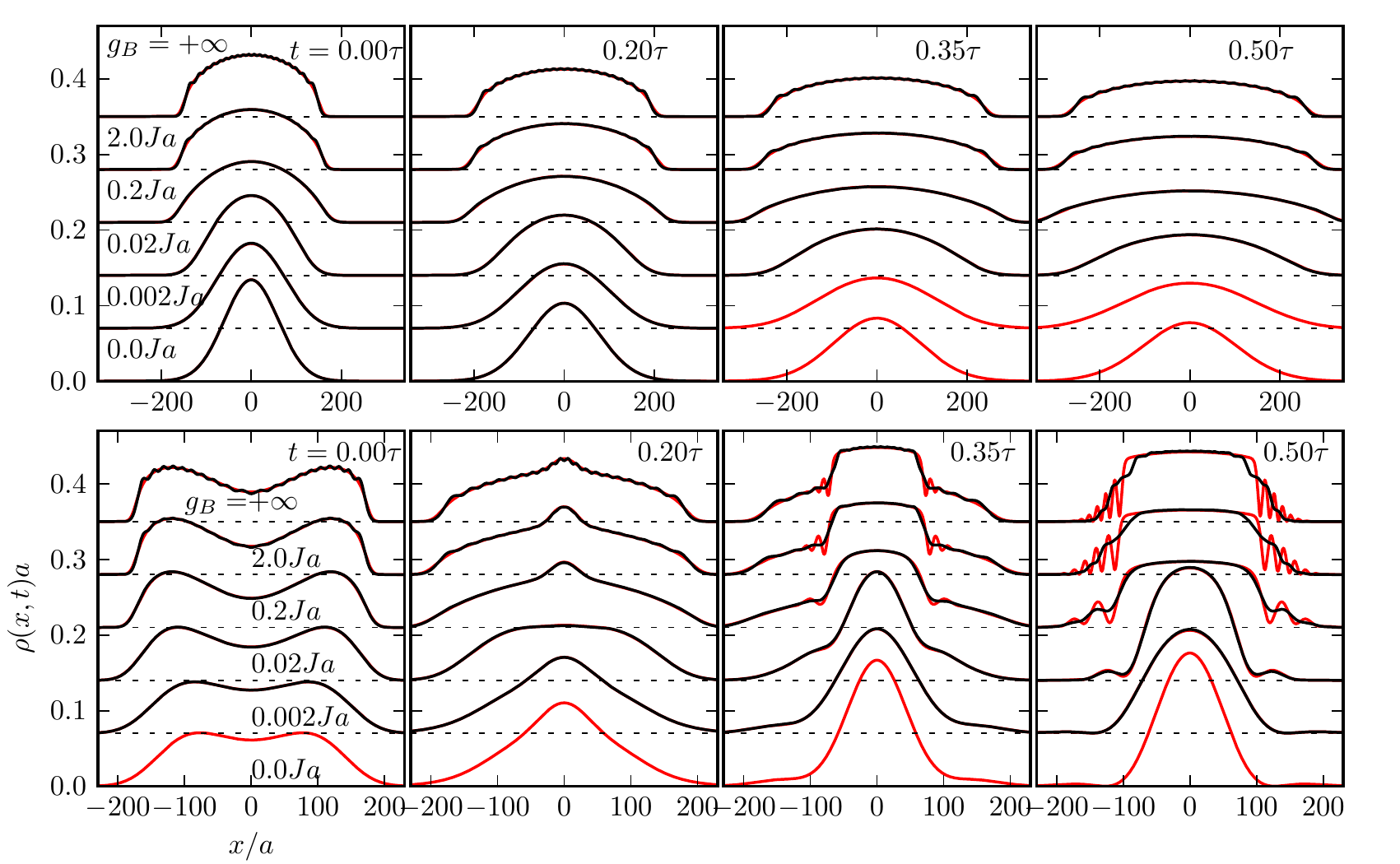}
\caption{\label{fig1} (Color online) Comparison between TDMRG data (black line) and the solutions of the hydrodynamic equation~(\ref{eq:gnlse}) (red line). The top row refers to the breathing mode quench while on the bottom one the collision of clouds is illustrated. Several different values of $g_B$ are shown, from free bosons $g_B = 0$ (bottom) up to the Tonks-Girardeau limit $g_B = +\infty$ (top). Density profiles for different interaction strength have been shifted vertically for clarity. Time is in unit of the half period defined as  $\tau = \pi/\omega_1$ for the collision quench and as $\tau = \sqrt{3}\pi/\omega_1$ for the breathing quench where the value of $\omega_1$ is given in Table~\ref{tb:frequencies}.  In the case of the breathing mode the perfect match between TDMRG and hydrodynamic (the density profiles are overlapping) is a strong indicator of the accuracy of the simulations. In the case of the collision quench, shock waves formation is signalled by steep density profiles at $t\sim 0.35\tau$ and the growth of oscillations present only in the hydrodynamic profiles. The amplitude of these oscillations increases with $g_B$ and they are very evident for $g_B= 2.0Ja$ and in the Tonks-Girardeau limit at $t=0.5\tau$. This leads to the breakdown of the hydrodynamic approximation Eq.~(\ref{eq:gnlse}) which is unable to capture the true quantum dynamics where the ripples are smoothed out by quantum fluctuations.}
\end{figure*}

\subsection{\label{sec:hydrodynamic}Hydrodynamic description}
In this work we compare our TDMRG results with the best possible (to our knowledge) hydrodynamic description for the present case, namely a generalized nonlinear Schr\"odinger equation (GNLSE)~\cite{Dunjko:2001,Kim:2003,Lieb:2003,Salasnich:2004,
Damski:2004,Salasnich:2005,Damski:2006}
%
%
\begin{equation}\label{eq:gnlse}
i\hbar \partial_t\Psi(x,t) = \left[-\frac{\hbar^2}{2m} \partial_x^2 + \phi(\rho) + V(x,t)\right]\Psi(x,t)\,,
\end{equation}
where $\Psi(x)$ is a complex field, $V(x,t)$ is specified by~(\ref{eq:post_pot})  and~(\ref{eq:init_pot}), $\rho(x) = |\Psi(x)|^2$ is the density, and $\phi(\rho)$ the Gibbs free energy per particle, or chemical potential, obtained from the Bethe Ansatz solution of the Lieb-Liniger model. In the Gross-Pitaevskii limit ($g_B\to 0$)  $\phi(\rho)=  g_B\rho$ while in the hard-core limit ($g_B\to +\infty$) $\phi(\rho)= \frac{\pi^2\hbar^2}{2m}\rho^2$. Accurate numerical values of $\phi(\rho)$ for intermediate interactions are available~\cite{Dunjko:2001}.

Eq.~(\ref{eq:gnlse}) can be written alternatively in a more standard hydrodynamic form by using the  {\it de Broglie ansatz} $\Psi(x,t) = \sqrt{\rho(x,t)} e^{iS(x,t)/\hbar}$ and separating the real and imaginary part. The result is the quantum Euler equations~\cite{Damski:2006}
\begin{gather}
\partial_t\rho = -\partial_x\left(\rho v\right)\,,\label{eq:cont}\\
\partial_t v + v\partial_x v = -\frac{1}{m}\partial_x\left(\phi(\rho) -\frac{\hbar^2}{2m}\frac{\partial_x^2\sqrt{\rho}}{\sqrt{\rho}}+V(x,t)\right)\,,\label{eq:mom}
\end{gather}
where the velocity field $v(x,t) = \partial_x S(x,t) /m $ has been introduced. Without the \textit{quantum pressure} $-(\hbar^2\partial_x^2\sqrt{\rho})/(2m\sqrt{\rho})$ Eqs.~(\ref{eq:cont}) and~(\ref{eq:mom}) amount to a simple Local Density Approximation (LDA), but this term needs to be included in order to reproduce the free bosons limit ($g_B\to 0$). Note that there are no free parameters in Eq.~(\ref{eq:gnlse}) or equivalently in Eqs.~(\ref{eq:cont}) and~(\ref{eq:mom}).
The GNLSE Eq.~(\ref{eq:gnlse}) has been solved numerically using a time-splitting spectral method~\cite{tssm}. We used a fourth-order Trotter expansion to perform imaginary time evolution in the initial potential (\ref{eq:init_pot}), thus providing the initial state $\Psi(x,t = 0)$. A sixth order Trotter expansion was used to evolve the system in the quenched potential (\ref{eq:post_pot}), the same expansion employed for TDMRG.

\begin{figure}
\includegraphics{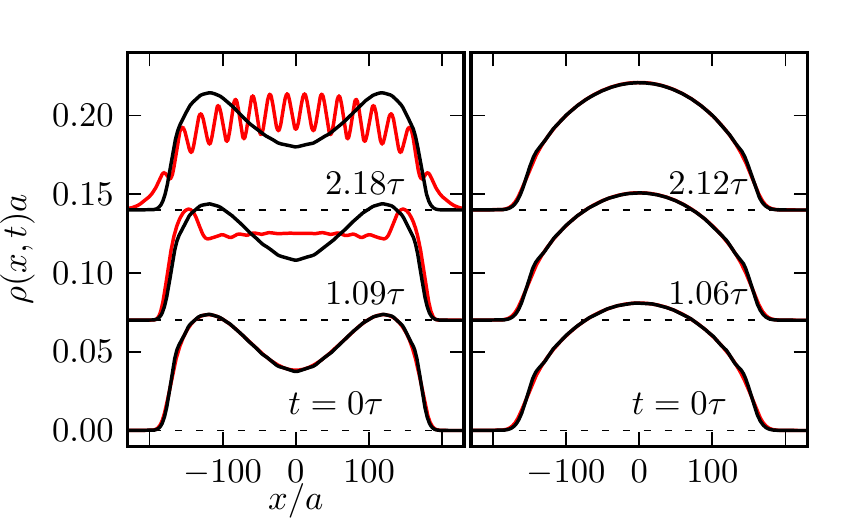}
\caption{\label{fig2} (Color online) Comparison between TDMRG (black line) and hydrodynamic (red line) density profiles at $t=0$ (bottom), after one oscillation period $\tau^*(g_B)$ (middle) and after two periods (top). The density profiles at different times have been shifted in the vertical direction.
How the renormalized oscillation period has been extracted from the TDMRG data is explained in Sec.~\ref{sec:period} and Fig.~\ref{fig:three}. Note that the hydrodynamic simulations match the TDMRG results for several oscillations periods in the case of the breathing mode while in the collision of clouds the approximation breaks down before a single oscillation is completed due to shock waves formation (see Fig.~\ref{fig1}). In both cases the full quantum dynamics exhibit (quasi-)periodicity for any interaction strength (the data shown here refers to $g_B=1.0Ja$).}
\end{figure}

\section{\label{sec:shock_waves} Classical and Quantum Hydrodynamics}

\subsection{Shock waves}
TDRMG and hydrodynamics are compared in Fig.~\ref{fig1} both for the breathing mode quench and the clouds' collision for different values of the interaction strength ($g_B = 0,\dots,+\infty$). In the case of the breathing mode (upper panels) the excellent agreement at all times -- even longer than those shown in Fig.~\ref{fig1} (see Fig.~\ref{fig2} and below) -- and for all $g_B$'s indicates that lattice effects are negligible (continuum limit)~\cite{Peotta:2011,Peotta:2012} and, quite surprisingly, the hydrodynamic description works well for just $N= 20$ particles. An analogous rapid crossover from a few particles to a many particles regime has been observed in Ref.~\onlinecite{Peotta:2011}. A feature that hydrodynamics is unable to capture are the small oscillations  in the density profile visible in the TDMRG data for any $t \geq 0$ in the strongly interacting limit. These are called \textit{shell effects}~\cite{Gleisberg:2000,Vignolo:2000,Wonneberger:2001,
Brack:2001,Wang:2002,Mueller:2004} and are a feature of the ground state that persists during the evolution. We stress that the agreement between the results obtained with two completely different methods such as TDMRG and hydrodynamics is a strong check of the accuracy of our simulations.

Contrary to the breathing mode, the hydrodynamic description can capture the dynamics of colliding clouds, shown in the lower panels of  Fig.~\ref{fig1}, only up to a time $t\sim 0.35\tau$ when oscillations form in the GNLSE solutions, corresponding to the formation of shock waves (\textit{gradient catastrophe}~\cite{Kulkarni:2012,Whitham:book}). The oscillation amplitude increases with the interaction strength and is maximal in the Tonks-Girardeau limit. These shock waves with oscillatory behaviour are known as \textit{dispersive} and occur in inviscid fluids, e.g., Bose-Einstein condensates~\cite{Hoefer:2006,Meppelink:2009,Kulkarni:2012,Lowman:2013}.

Our TDMRG results are very similar to the experimental data reported in Ref.~\onlinecite{Joseph:2011} where viscosity was introduced in the hydrodynamic equations to describe shock waves, while in the TD-DFT calculation in Ref.~\onlinecite{Ancilotto:2012} a renormalized kinetic term $\lambda\partial_x^2\Psi$ was used for the same reason. This is not justified here since Eq.~(\ref{eq:gnlse}) has no free parameters and it is an excellent approximation up to the gradient catastrophe for \textit{any} $g_B$. Introducing viscosity would contradict the fact that almost no dissipation is present in our system as we will show below. It is however unclear what kind of dispersive term should be used in our case to reproduce the exact quantum dynamics where the oscillations are suppressed with respect to the GNLSE dynamics. A discussion of the dissipative or dispersive nature of shock waves in quantum gases can be found in Refs.~\onlinecite{Ancilotto:2012,Lowman:2013}. 

As it is nicely illustrated in Fig.~\ref{fig1} the dynamics of these quantum shock waves for finite $g_B$ are in fact continuously connected to the Tonks-Girardeau limit (free fermion, $g_B \to +\infty$) a fact anticipated in Ref.~\onlinecite{Bettelheim:2012}. Suprisingly enough the hydrodynamics of free fermions is still poorly understood and has been the subject of recent works~\cite{Bettelheim:2012,Mirlin:2012}.

\begin{figure}
\includegraphics{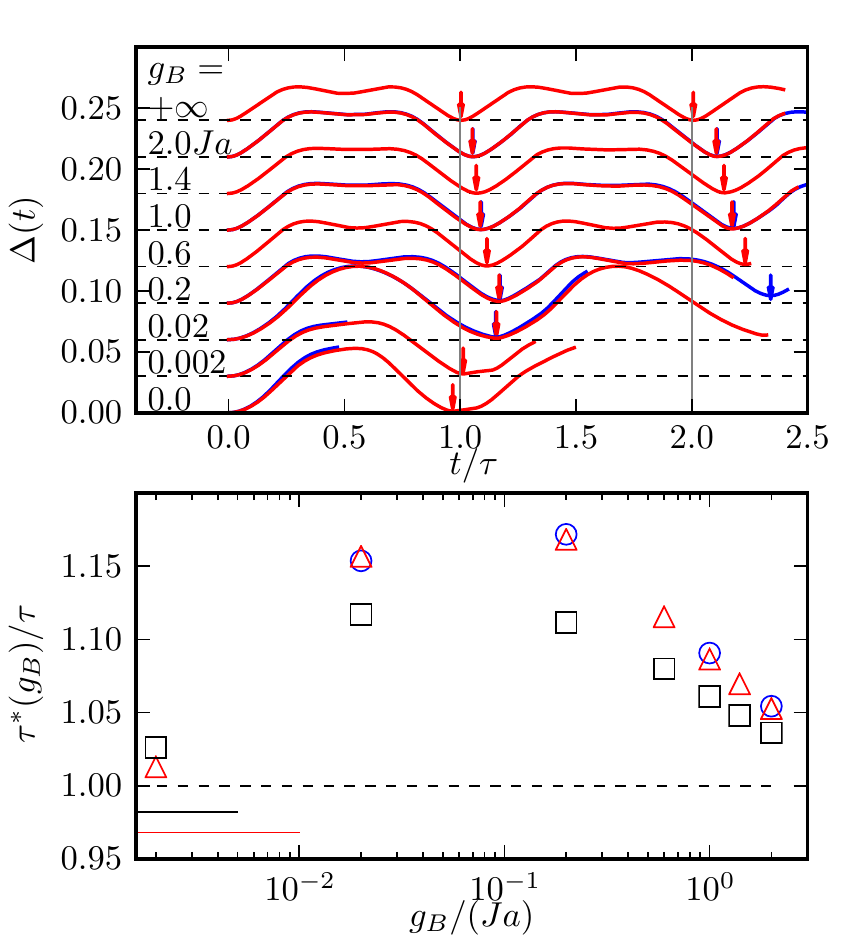}
\caption{\label{fig:three} (Color online) Upper plot, deviation $\Delta(t)$ as a function of time [Eq.~(\ref{eq:delta})]. In blue are the data relative to the Bose-Hubbard discretization and in red the ones relative to the XXZ spin chain discretization. The arrows indicate the instants where the density is closest to the initial one (minima of $\Delta(t)$). In the lower plot the renormalized oscillation period $\tau^*(g_B)/\tau$ is shown as a function of the interaction strength $g_B$ extracted from the minima of $\Delta(t)$ (see the upper plot). The red triangles (blue circles) refer to the XXZ  (Bose-Hubbard) discretization. The black squares are relative to the breathing period for which hydrodynamics and TDMRG agree. The red (black) line at the bottom left represents the frequency extracted from the $g_B =0$ data for the collision (breathing) quench. They deviate from the exact result $\tau^*(0) = \tau$ in the continuum limit since lattice effects distort the density profile in time. As can be seen in Fig.~\ref{fig1} higher densities are explored in the $g_B=0$ case thus the continuum limit approximation is less accurate.}
\end{figure}

\subsection{\label{sec:period}Oscillation frequency shift}
Although the system has a strongly nonequilibrium and nonlinear dynamics we find from the TDMRG simulations that the initial density profile and thus the initial state are recovered after a time of order $\tau$ for any $g_B \in [0,+\infty]$ both in the breathing and in the collision quenches. This remarkable recurrence, that would be expected only in the limit of small oscillations, is shown in Fig.~\ref{fig2} for $g_B = 1.0Ja$.
The hydrodynamics results for the breathing quench (right panel of Fig.~\ref{fig2}) show a remarkable agreement with the quantum dynamics for times at least as long as few oscillation periods, while they deviate rapidly in the collision quench (left panel).

The profiles in Fig.~\ref{fig2} are shown at times $t = 0,\,\tau^*(g_B),\,2\tau^*(g_B)$
where the renormalized oscillation period $\tau^*(g_B)$ has been extracted as follows.
We use  the mean square deviation of the density profile at time $t$ from the initial one
\begin{equation}\label{eq:delta}
\Delta(t) = \frac{1}{N}\sqrt{\int dx\,(\rho(x,t)-\rho(x,0))^2}\,,
\end{equation}
shown in the upper panel of Fig.~\ref{fig:three} for the collision quench.
These curves have been obtained from TDMRG data since hydrodynamics is unreliable in this case. Moreover we have compared results using the two different discretizations employed~[App.~\ref{appendix:discretization}] and found no significant differences. $\Delta(t)$ essentially drops to zero at times $t \sim\tau$ and $t \sim 2\tau$ indicating that the system has approximately returned to the initial state. The times at which the first minimum occurs is precisely $\tau^*(g_B)$. The second minimum occurs at $2\tau^*(g_B)$ to a good approximation.
The results for the renormalized period are shown in the lower panel of Fig.~\ref{fig:three}.
In the exactly solvable limits $g_B=0$ and $g_B=+\infty$ the period is not renormalized. In between these extrema it has a nonmonotonic behaviour with a maximum in the interval $0.02< g_B /(Ja) <0.2$ ($0.2 <\gamma < 2$). The almost perfect periodicity observed for any $g_B$ is a strong indication of very small dissipation, in agreement with experimental results~\cite{Kinoshita:2006}. The collision period is found to be measurably larger than the breathing period -- for which hydrodynamics is accurate~\cite{Astrakharchik:200} --, a fact that could be easily tested experimentally.

While the dynamics obtained from TDMRG are essentially periodic, the ones obtained from Eq.~(\ref{eq:gnlse}) are rather chaotic in the case of the collision quench. We remind the reader that the static density  and the dynamics up to the gradient catastrophe are well captured by the GNLSE~(\ref{eq:gnlse}) (see the left panels of  Fig.~\ref{fig1}).
  A simple explanation of this phenomenon is that in the case of the breathing mode the density is slowly varying in space and what counts is just the local pressure in the LDA sense, which is reproduced by Eq.~\ref{eq:gnlse}, or by Eqs.~\ref{eq:cont} and~\ref{eq:mom}, by definition. However in the case of the collision quench where shock waves are formed, gradient corrections on top of LDA become crucial and the quantum pressure is inadequate since it leads to a qualitatively different evolution.
A different point of view is that Eq.~\ref{eq:gnlse} with the external potential set to zero entails the conservation only of particle number, momentum and energy and produces a chaotic dynamics, while the full quantum dynamics is subject to an infinite number of conservation laws or, in other words, the Lieb-Liniger Hamiltonian is integrable~\cite{Sutherland_book}. It appears that the integrability breaking due to the external potential is small in this case and slightly affecting the quantum dynamics.

\begin{figure*}
\includegraphics{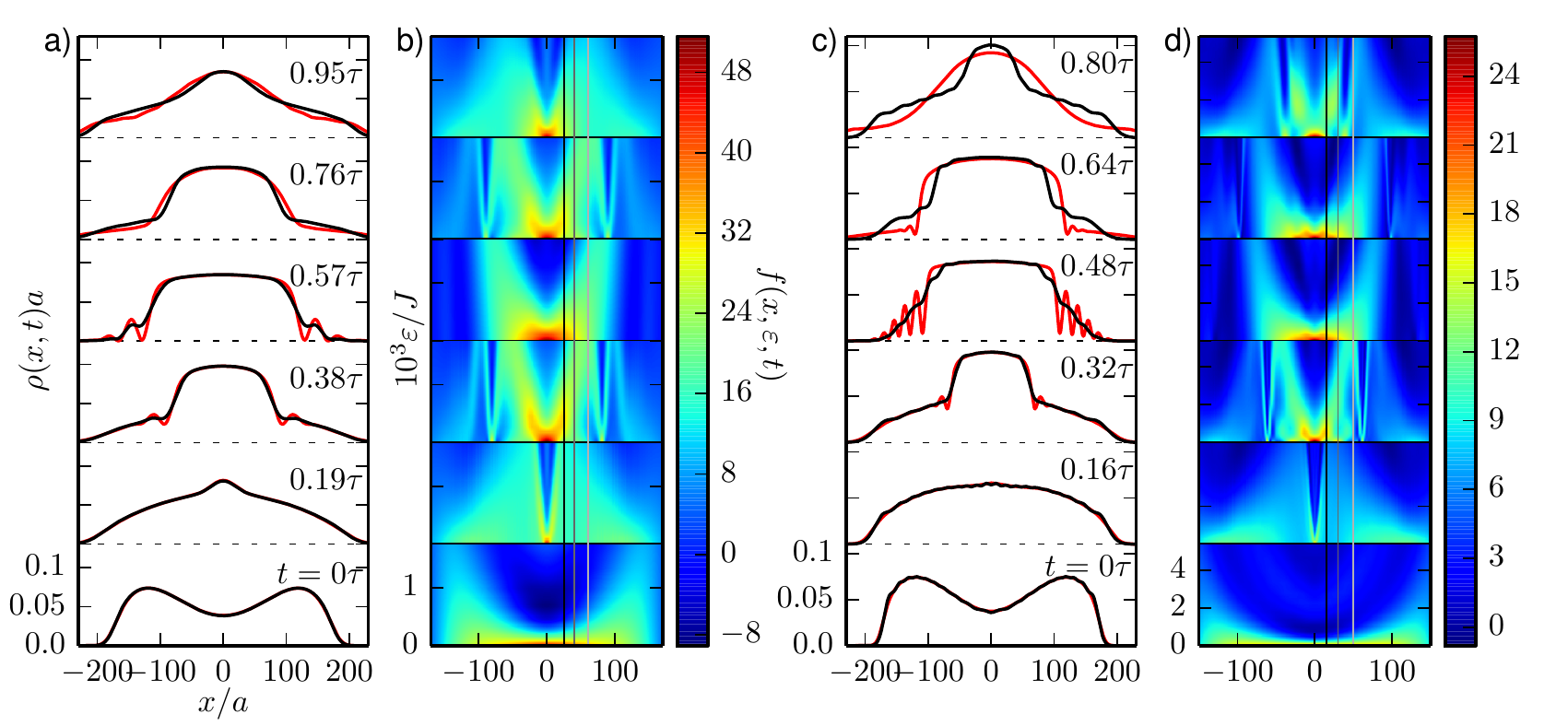}
\caption{\label{fig4} (Color online) \textbf{a)} Snapshots of the density profile $\rho(x,t)$ for the clouds' collision as in Fig.~\ref{fig1} ($g_B = 0.2 Ja$, $\gamma = 2$). \textbf{b)} Energy distribution function $f(x,\varepsilon,t)$~(\ref{eq:erg_dist}) calculated from the Wigner function (\ref{eq:wigner}). The color plots show the values of $f(x,\varepsilon,t)$ in the $(x,\varepsilon)$ plane at the corresponding times in panel a). The vertical black, grey and light grey lines correspond to the sections for fixed $x$ of $f(x,\varepsilon,t)$ shown in Fig.~\ref{fig5}. Note that shock-wave formation is characterized by a highly nonequilibrium distribution. \textbf{c)-d)} Same as panels a) and b) but for $g_B = 2.0Ja$ $(\gamma = 20)$.}
\end{figure*}
\begin{figure}
\includegraphics{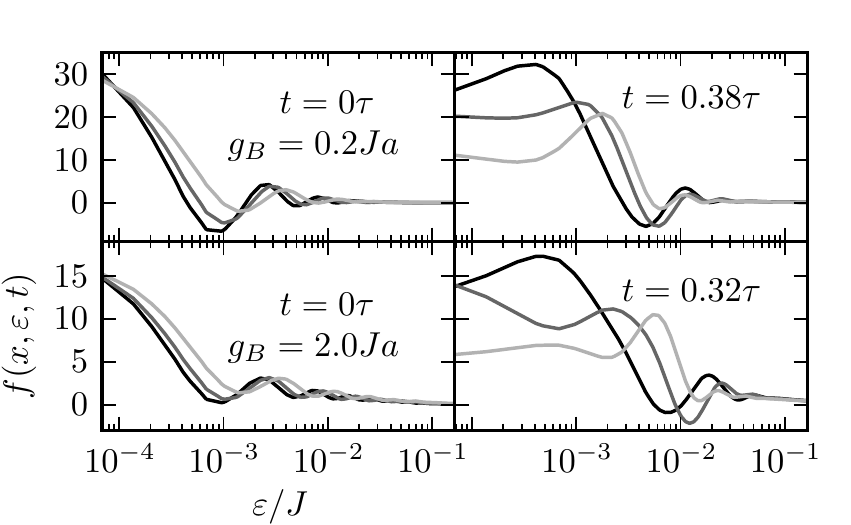}
\caption{\label{fig5} Energy distribution function $f(x,\varepsilon,t)$ corresponding to first ($t = 0\tau$) and third ($t \sim 0.3\tau$)  snapshots in Fig.~\ref{fig4}, panels \textbf{b)} and \textbf{d)}. The black, grey and light grey lines correspond to $x/a = 25, 40, 60$ for the two upper quadrants ($g_B = 0.2Ja$) [Fig.~\ref{fig4} panel \textbf{b)}], and $x/a = 15, 30, 50$ for the two lower quadrants ($g_B = 2.0Ja$) [Fig.~\ref{fig4} panel \textbf{d)}], respectively. Note that, neglecting oscillations for large $\varepsilon$ -- due to the finite number of particles -- the initial distribution decreases monotonically, while a maximum develops for finite $\varepsilon$ after the shock-wave formation at $t\sim 0.3\tau$, \textit{i.e.} a population inversion.}
\end{figure}
\section{Population inversion}
In order to study in more detail the shock wave dynamics we use the Wigner function~\cite{Bettelheim:2012,Mirlin:2012}
%
%
\begin{equation}\label{eq:wigner}
W(x,p,t) = \frac{1}{\hbar \pi} \int dy\,
\rho(x+y,x-y;t)
e^{\frac{2ipy}{\hbar}}\,,
\end{equation}
where $\rho(x',x;t)=\langle \hat \Psi^\dagger\left(x,t\right)\hat \Psi\left(x',t\right)\rangle$ is the one-body density matrix. The one-body density matrix can be easily extracted by contracting the wavefunction in MPS form~\cite{tdmrg}.
\begin{figure}
\includegraphics{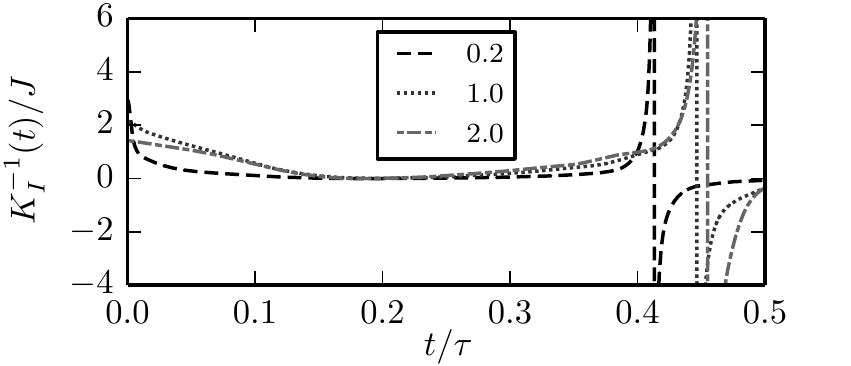}
\caption{\label{fig6} Inverse information compressibility as a function of time for various interaction strengths $g_{B}/(Ja) = 0.2,\,1.0,\,2.0$. Note the divergence in correspondence to a fully developed shock wave at $t = 0.4\div0.5\tau$. See Fig.~\ref{fig4}, snapshots at $t=0.38\tau$ in panel a) and at time $t=0.32\tau$ in panel c).}
\end{figure}
Neglecting negative values, $W(x,p,t)$ can be thought of as a \textit{local momentum ($p$) distribution} as in the Boltzmann equation. Oscillations and negative values of the Wigner function obviously spoil its interpretation as a local momentum distribution. However, we have found in the case of free fermions, where a scaling with the number of particles is possible, that such features do not preclude a well defined Fermi step with increasing $N$.

The local \textit{energy} distribution $f(x,\varepsilon,t)$ is defined with respect to the local co-moving (Lagrangian) reference frame with velocity $mv(x,t) =j(x,t)/\rho(x,t)$~\cite{mybook}, with $j(x,t) = \int dp\,pW(x,p,t)$ and $\rho(x,t) = \int dp\, W(x,p,t)$. Thus $\int dp\, W(x,p-mv(x),t) = 0$ and the energy distribution reads
%
%
\begin{equation}\label{eq:erg_dist}
f(x,\varepsilon,t) = 2\pi \hbar \sum_{s = \pm}W(x,s\sqrt{2m\varepsilon}-mv(x))\,.
\end{equation}
This is the quantity shown in the color plots in Fig.~\ref{fig4} and for selected values of the position $x$ in Fig.~\ref{fig5}. At $t = 0$ [bottom of Fig.~\ref{fig4} and left panels of Fig.~\ref{fig5}] the distribution $f(x,\varepsilon,t)$ decreases monotonically  with $\varepsilon$ (leaving aside oscillations related to the finite particle number) indicating an equilibrium energy distribution.

In correspondence of the shock wave formation ($t \sim 0.3 \tau$) the energy distribution is no longer an equilibrium one, $f(x,\varepsilon,t)$ is larger for values of $\varepsilon$ away from zero [right panels of Fig.~\ref{fig5}], signalling a population inversion, namely an effective negative temperature. Population inversion leads to the break down of the LDA and to the deviation from the GNLSE (Eq.~\ref{eq:gnlse}) solution. The energy distribution function in the usual sense is $\sqrt{m/(2\varepsilon)}f(x,\varepsilon,t)$ and  contains the 1D density of states factor $\sqrt{m/(2\varepsilon)}$. However, for our purposes the definition in Eq.~(\ref{eq:erg_dist}) is more appropriate since for a classical system at equilibrium $f(x,\varepsilon,t) \propto e^{-\beta \varepsilon}$ and a monotonically increasing behaviour in the distribution directly corresponds to a negative temperature. This would not be immediately apparent if the 1D density of states had been taken into account.

\subsection{Negative temperature}
In order to corroborate the presence of a negative temperature out of equilibrium we characterize the state of the system with the \textit{information compressibility}~\cite{DiVentra:2009}, which measures the relative change of the number of available microstates of an open system in response to an energy variation. In our case we deal with a closed and finite quantum system. However, it is always possible to trace out part of the system and study the remaining half as an open one. In our case we focus on the reduced density matrix of half of the chain
\begin{equation}\label{DM}
\begin{split}
\hat \rho_{\rm half} (t) = &\sum_{n_{L/2+1},\ldots,n_{L}}\braket{n_1,\ldots,n_{L/2},n_{L/2+1},\ldots, n_L}{\Psi} \\ &\times\braket{\Psi}{n_1,\ldots,n_{L/2},n_{L/2+1},\ldots,n_L}\,.
\end{split}
\end{equation}
The concept of information compressibility has been introduced in Ref.~\onlinecite{DiVentra:2009} as a mean to characterize out of equilibrium  states of open systems.
Expectation values can be easily extracted from $\rho_{\rm half}(t)$ if the state $\braket{\{n_i\}}{\Psi}$ is in MPS form~\cite{tdmrg}.
Call $\Omega$ the number of microstates available to the system. The information compressibility is then defined as the relative variation of the number of microstates with respect to the energy variation at time $t$~\cite{DiVentra:2009}
\begin{equation}
K_I(t) = \left.\frac{1}{\Omega}\frac{\delta \Omega}{\delta E}\right|_{t}\,.
\end{equation}
Using the microcanonical relation $\Omega = \exp\left(S/k_\text{\tiny B}\right)$ one arrives at the computationally more convenient definition
\begin{equation}\label{eq}
K_I(t) = \left.\frac{1}{k_\text{\rm B}} \frac{\partial S}{\partial t}\frac{\partial t}{\partial E}\right|_t\,.
\end{equation}
Note the similarity of this quantity with the thermodynamic definition of inverse temperature~\cite{DiVentra:2009}.
Given a system with density matrix $\hat \rho(t)$ and Hamiltonian $\mathcal{\hat H}$, the energy is $E(t) = \mathrm{Tr}[\hat \rho(t)\mathcal{\hat H}]$ and the thermodynamic entropy $S(t) = -k_\text{\tiny B}\mathrm{Tr}\left[\hat \rho(t)\ln\hat \rho(t)\right]$.
We used for $\hat \rho(t)$ the density matrix  of half of the system $\hat \rho_{\rm half}(t)$ defined above and $\mathcal{\hat H} = \mathcal{\hat H}_{\rm internal}$ is the part of the Hamiltonian relative to the \textit{internal energy} of the system, namely the sum of kinetic and interaction energy but excluding the potential energy due to external forces
\begin{equation}
\mathcal{\hat H}_{\rm internal} = \int dx\, \bigg[\frac{\hbar^2}{2m}|\partial_x \hat \Psi(x)|^2 + \frac{g_B}{2}|\hat \Psi(x))|^4\bigg]\,.
\end{equation}
This is consistent with the thermodynamic definition of inverse temperature as the derivative of the entropy with respect to the internal energy.

The inverse information compressibility is shown in Fig.~\ref{fig6} for various interaction strengths. The interesting point is the divergence of $K_I^{-1}(t)$ in all cases for $t/\tau \sim 0.4$, \textit{i.e.}, in correspondence to a fully-developed shock wave, and negative values of this quantity at later times. If we interpret the inverse compressibility as an effective temperature, this behaviour is clearly suggestive of a population inversion, in agreement with our previous results.

\section{Conclusions and perspectives}

In this work we have used TDMRG, an essentially exact method, and an approximate hydrodynamic description based on the GNLSE to study numerically the debated \qql quantum Newton cradle\qqr experiment~\cite{Kinoshita:2006}. We find that, when the two clouds of atoms collide, shock waves occur almost immediately after the quench, a fact which has been previously overlooked, suggesting interesting connections with the growing literature on the subject of shock waves in ultracold gases~\cite{Hoefer:2006,Meppelink:2009,Joseph:2011,Bulgac:2012,
Salasnich:2012,Ancilotto:2012,Kulkarni:2012,Lowman:2013}. On the contrary, in a quench where the breathing mode is excited shock waves are absent. Interestingly, while shock waves greatly affect the GNLSE dynamics by triggering an aperiodic and chaotic behaviour, this does not occur in the full quantum dynamics where the system is found to return to the initial state after (approximately) half of the harmonic trap period, as in the case of the breathing period. We observe essentially no decay within the time scales that we have been able to explore ($\lesssim 3\tau$), an indication of the extremely small dissipation in the system. 
This suggest that the shock structure is controlled by a dispersive term which is however rather different from the quantum pressure in Eq.~(\ref{eq:mom}), since it leads to a qualitatively different dynamics and to an oscillatory structure different from the usual one~\cite{Whitham:book,Lowman:2013,Mirlin:2012,Bettelheim:2012}. We provide results for the oscillation period as a function of the interaction strength in the case of the collision of clouds, a nonperturbative result that, to our knowledge, can be obtained with no method other than TDMRG.

The experiment in Ref.~\onlinecite{Kinoshita:2006} is important for the problem of thermalization, namely what is the appropriate statistical ensemble that can describe the state asymptotically reached by an integrable system. Although this is not our main focus here, we point out that TDMRG could be useful in the future for this purpose since in the kind of quench that we study the entanglement has very little or no growth, which implies that the computational cost grows linearly with the maximum time reached in a simulation [see Ref.~\cite{SOM} and App.~\ref{appendix:tdmrg}]. In the present case we study only the first few oscillations. Notice, however, that in a three-dimensional collision of two Bose-Einstein condensates the thermalization scale is $\lesssim 2\tau$~\cite{Kinoshita:2006}, well within reach of our method. Importantly, we observed that the dynamics for finite interaction strength $g_B$ is continuously connected~\cite{Bettelheim:2012} to the Tonks-Girardeu case (free fermions, $g_B = +\infty$) for which the dynamics in a harmonic trap is stricly periodic, \textit{i.e.} there is \textit{no decay} towards a stationary asymptotic state. Understanding if this is the case also for arbitrary finite interactions is an important question, and in the future it may be possible to provide some lower bounds on the decay rate using TDMRG.

The results presented here are also relevant to the broad problem of understanding the hydrodynamics of quantum gases, namely to provide an effective description using as dynamical variables only the observables of interest, such as density and velocity fields~\cite{mybook}.
Such a description can be of great value since it is computationally more affordable than full quantum simulations such as those provided by TDMRG.
This is the same point of view adopted by TD-DFT in its orbital free formulation~\cite{Ligneres:2005}. In fact the Runge-Gross theorem~\cite{Gross:1984} of TD-DFT guarantees that an exact hydrodynamic description of quantum dynamics exists~\cite{mybook}, although the analytical expression of the stress tensor is unknown even for free fermions~\cite{Bettelheim:2012,Mirlin:2012}.
The use of DMRG to study Density Functional Theory in an exact setting, has been put forward in Ref.~\onlinecite{White:2012} in the context of ground-state calculations. Here, we have approached the dynamical problem for one of the simplest many-body systems in the same fashion. We emphasise that a better understanding of shock waves even in the Tonks-Girardeau limit is an important step towards the goal of a better hydrodynamic description of ultracold gases.

Finally, we have shown that quantum shock waves lead to a population inversion in the local energy distribution, namely to a negative effective temperature, a result confirmed by a possible definition of temperature out of equilibrium put forward in Ref.~\onlinecite{DiVentra:2009}. Our results suggest that statistical ensembles with negative temperatures for the motional degrees of freedom, as shown in Ref.~\onlinecite{Braun:2013}, are a common feature in collision experiments with ultracold gases~\cite{Kinoshita:2006}.

\begin{acknowledgments}
This work has been supported by DOE under Grant No. DE-FG02-05ER46204. We thank C.-C. Chien for a critical reading of
our paper and B. Damski and L. Glazman for useful suggestions. The numerical results presented in this work have been obtained by using an implementation of the TDMRG code with Matrix Product States, developed by the team coordinated by Davide Rossini at the Scuola Normale Superiore, Pisa (Italy).
\end{acknowledgments}

\appendix

\section{TDMRG simulations}\label{appendix:tdmrg}

In the TDMRG simulations a Matrix Product State (MPS) representation~\cite{tdmrg} of the wavefunction has been employed
\begin{equation}\label{eq:mps}
\braket{n_1,n_2,\ldots,n_{L-1},n_{L}}{\Psi} = \bm{A}^{[n_1]}\cdot \bm{A}^{[n_2]}\cdot \ldots\cdot \bm{A}^{[n_{L-1}]}\cdot \bm{A}^{[n_L]}\,,
\end{equation}
with $\{n_i\}_{i = 1,\ldots,L}$ a given set of occupancies of the lattice with length $L$. The matrix $\bm{A}^{[n_i]}$ for fixed $n_i$ has dimension $m_{i-1}\times m_i$ where $m_i$ is called the \textit{link  dimension}, an integer number attached to the link connecting site $i$ and site $i+1$.  The link dimension is position dependent and it is the crucial parameter that needs to be tuned to find a balance between accuracy and speed~\cite{tdmrg}. For open boundary conditions $m_0 = m_L = 1$. The dot \qql $\,\cdot\,$\qqr denotes the matrix multiplication.

A standard trick for increasing the speed of TDRMG is the use of the conservation of the number of particles ($\sum_i n_i = N$). This leads to a block structure for the matrices $\bm{A}^{[n_i]}$. It can be easily checked that large blocks of $\bm{A}^{[n_i]}$ are zero and the size of the MPS is greatly reduced.

During the time evolution the link dimension $m_i$ is kept to a low value by performing a singular value decomposition~\cite{tdmrg} (SVD) of a two-site matrix $\bm{M}^{[n_in_{i+1}]}_{\ell_{i-1}\ell_{i+1}} = \sum_{\ell_i}\bm{A}^{[n_i]}_{\ell_{i-1}\ell_i}\bm{A}^{[n_{i+1}]}_{\ell_i\ell_{i+1}}$ and discarding the lowest singular values compatibly with the condition
\begin{equation}
 \epsilon > \sum_{\rm discarded\; \sigma_{\ell_{i}}}\sigma_{\ell_i}^2\,,
\end{equation}
where  $\sigma_{1} \geq \sigma_{2} \geq \dots \geq \sigma_{m_i-1} \geq \sigma_{m_i}$ are the singular values obtained by SVD and $\epsilon$ is the \textit{discarded weight}, a  small parameter that controls the precision. Note that according to this truncation procedure the link dimension $m_i$ adapts automatically in space and time to the evolving wavefunction of an inhomogeneous and out of equilibrium system.
The block structure carries over to $\bm{M}^{[n_in_{i+1}]}_{\ell_{i-1}\ell_{i+1}}$ and the SVD can be performed blockwise with considerable speed-up~\cite{tdmrg}. A MPS with a block structure imposed by the conservation of the number of particles and  the truncation prescription described above are the two ingredients that enable to simulate reliably the quench protocol presented in the main text for long enough times to observe several collisions of the clouds.  The same techniques have been employed successfully for the Fermi-Hubbard model in Ref.~\onlinecite{Peotta:2011} and for the Bose-Hubbard model with two species in Ref.~\onlinecite{Peotta:2012}. Additional details on the structure imposed on the MPS by particle number conservation can be found in Ref.~\onlinecite{Peotta:2013}.

In our simulations we used a discarded weight $\epsilon = 10^{-10}$ and we employed a sixth order Trotter expansion for the time evolution~\cite{trotter,Peotta:2011,Peotta:2012} with time step $\Delta t = 0.1\hbar/J$ for the BH discretization and $\Delta t = 0.05\hbar/J$ for the XXZ discretization. The reason for using a sixth order expansion has been discussed in Ref.~\onlinecite{Peotta:2011}.
The reliability of our simulations has been controlled in several ways. First, we checked our results against exactly solvable cases, namely free bosons and free fermions (hard-core limit $g_B\to +\infty$) verifying that the exact diagonalization results are indistinguishable from the TDMRG ones over several oscillation half periods $\tau = \pi/\omega_1(t>0)$, the scale of one collision.
Second, the discarded weight has been lowered to $\epsilon = 10^{-11}$ without observing significant differences in the evolved density profile $\rho(x,t)$.
Finally the comparison between hydrodynamic and TDMRG data in the case of the breathing mode quench [see Fig.~\ref{fig1} and~\ref{fig2}] is by itself an unbiased check, for any value of the interaction strength, of the accuracy of our simulations over several oscillation periods.

In the Supplementary Online Materials~\cite{SOM} we provide animations of the density profiles obtained both with TDMRG and GNLSE, alongside the corresponding link dimension $m_i$ and block entropy $S_i$~\cite{tdmrg} illustrating the important point that in our TDMRG simulation the entanglement growth is not so dramatic, which allows in principle to reach longer times than those presented here.

\section{Discretization of the Lieb-Liniger Hamiltonian}\label{appendix:discretization}
The Lieb-Liniger model has been discretized in two distinct ways:
\begin{itemize}
\item \textbf{Bose-Hubbard discretization}~\cite{Kollath:2005,Peotta:2012,Knap:2013}
\be
\label{eq:bose-hubbard}
\begin{split}
{\hat {\cal H}}_{\rm Bose-Hubbard} = &-J\sum_i ({\hat
  b}^\dagger_{i}{\hat b}_{i+1}+ {\rm H.c.}) \\ &+ \frac{U}{2}\sum_i {\hat n}^2_{i}
+ \sum_iV_i{\hat n}_i\,,
\end{split}
\ee
with $\hat b_i,\,\hat b_i^\dagger$ bosonic annihilation and creation operators on the $i$-th site and ${\hat n}_i = \hat b_i^\dagger \hat b_i$ the corresponding on-site density operator. The maximal occupacy as been truncated to $n_i \leq 6$ in the simulations.

\item \textbf{XXZ spin chain discretization}~\cite{Muth:2010,Muth:2010bis}
\begin{equation}\label{eq:XXZ}
\begin{split}
\mathcal{\hat H}_{\rm XXZ} = &-J\sum_i(\hat c_i^\dagger\hat c_{i+1} + {\rm H.c.}) \\&-\frac{2J}{1+U/(4J)}\sum_i\hat n_i\hat n_{i+1} +\sum_iV_i\hat n_i\,,
\end{split}
\end{equation}
where $\hat c_i$,$\hat c_i^\dagger$ are fermionic annihilation and creation operators  and $\hat n_i = \hat c_i^\dagger\hat c_i$ is the on-site density operator. The Hamiltonian (\ref{eq:XXZ}) is equivalent to a XXZ spin chain after a Jordan-Wigner transformation. The equivalence between the Lieb-Liniger model and the low density limit of Hamiltonian~(\ref{eq:XXZ}) is a consequence of the Bose-Fermi mapping in 1D for \textit{arbitrary} $g_B$  discussed in Refs.~\onlinecite{Cheon:1998,Cheon:1999}. The discretization of the Hamiltonian for $p$-wave interacting  fermions is carried out in Ref.~\onlinecite{Muth:2010}.

\end{itemize}

The couplings in the above Hamiltonians are related to the continuum model as $J = \hbar^2/(2ma^2)$, $U = g_B/a$ and $V_i = V(x = ia)$. In our simulations we used the lattice spacing $a$ as unit of length and the hopping energy $J$ as unit of energy.
For a system with density $\langle \hat n_i\rangle/a = \rho$ the
dimensionless Lieb-Liniger parameter~\cite{Lieb:1963a,Lieb:1963b} reads
\begin{equation}\label{eq:ll_par}
\gamma = \frac{mg_B}{\hbar^2 \rho} = \frac{U}{2J\langle \hat n_i\rangle}\,.
\end{equation}

\end{document}